\documentclass[11pt]{llncs}
\usepackage{graphicx}
\title{Providing Virtual Execution Environments:\\ A Twofold Illustration.}
\subtitle{\sc{openlab Internal, May 2007}}
\author{Xavier Gr\'ehant \and J.M. Dana}
\institute{CERN openlab, Geneva, Switzerland}
\begin{document}
\maketitle
\begin{abstract}
Platform virtualization helps solving major grid computing challenges: share resource with flexible, user-controlled and custom execution environments and in the meanwhile, isolate failures and malicious code. Grid resource management tools will evolve to embrace support for virtual resource.

We present two open source projects that transparently supply virtual execution environments. Tycoon has been developed at HP Labs to optimise resource usage in creating an economy where users bid to access virtual machines and compete for CPU cycles. SmartDomains provides a peer-to-peer layer that automates virtual machines deployment using a description language and deployment engine from HP Labs.
These projects demonstrate both client-server and peer-to-peer approaches to virtual resource management. The first case makes extensive use of virtual machines features for dynamic resource allocation. The second translates virtual machines capabilities into a sophisticated language where resource management components can be plugged in configurations and architectures defined at deployment time.

We propose to share our experience at CERN openlab developing SmartDomains and deploying Tycoon to give an illustrative introduction to emerging research in virtual resource management.
\end{abstract}
\section{Introduction}
Platform virtualization systems are evolving so fast that scientific production grids have not yet caught up with them to improve the management of their resources. However research projects are going on in experimental grids and already cleared interesting ways.\\

We present Tycoon \cite{Lai:Huberman,Feldman:Lai}, a system developed mainly by Kevin Lai at HP Labs, that balances and optimizes resources while presenting virtual execution environments to remote users. Tycoon proposes an original mechanism to optimize resource usage on a global scale: a bidding mechanism and the architecture that makes it possible. Considering the important responsibility of platform virtualization in the novelty of Tycoon and the capabilities of its mechanism, as well as for other projects that bring up execution environments to remote users, we developed a system to help using virtualization in resource management: SmartDomains brings virtualization control to the level of SmartFrog, an advanced Java peer-to-peer configuration and deployment framework for distributed systems \cite{Goldsack:Guijarro,Sabharwal}. Both Tycoon and SmartFrog are projects developed at HP Labs. SmartDomains is developed at CERN openlab.\\

In section \ref{virtualization} we present how virtualization gives potential benefits to the area of resource management via execution environments supply. We illustrate it with the example of Tycoon detailed in section \ref{tycoon}. In section \ref{smartdomains} we present SmartDomains, and how it brings virtual machines management to a higher development level for convenience, where most fundamental mechanisms to deploy execution environments are already provided via SmartFrog.
\section{Benefits of virtualization to resource management} \label{virtualization}
\subsection{Virtualization}
Virtualization means creation of a computer resource abstraction. However the term is used broadly enough to cover any kind of resources (network, disk) and even services (in Service Oriented Architecture: SOA) or other concepts (virtual organizations). In this discussion, by virtualization we mean creation of virtual machines (VMs). Here again, the term ``virtual machine'' can be understood at several levels: hardware, operating system (OS) and application (like Java virtual machine). We talk about ``platform virtualization'' to name the emulation of virtual machines at the operating system level (it is an operating system which, in the end, is presented as an execution environment for our case of interest) potentially using hardware virtualization in the background. Again, there are some variants of platform virtualization (full, para-, in-kernel), the field as a whole is experiencing a tremendous growth, and most tools falling into this designation could theoretically have been used in the systems we describe.\\

Concretely, when virtualization is enabled, guest virtual machines can be booted from the host operating system. In the case of full virtualization like with VMWare\cite{Waldspurger} and KVM\cite{kvm}, an operating system kernel does not needs to be altered to run as a guest domain: it will not receive information from the host operating system nor from any other domain. With VMWare ESX, a software on the host operating system virtualizes resources (memory page tables, segment addresses, block devices, I/O interrupts, etc.) and generates code on the fly to replace priviledged instructions from the guest domain upon which x86 platforms do not behave conveniently to support virtualization. KVM's support for full virtualization is compiled in the kernel of the host operating system. By contrast, Xen\cite{Barham:Dragovic} developed para-virtualization, where guest domains apparent behavior is unchanged (application binary interfaces -ABIs- are unchanged) but is implemented differently: guest domains can access some information from the host OS such as IP address and time. To ensure security, their \emph{supervisor} runs at a less priviledged level than a virtualization system called the \emph{hypervisor} that runs on the host OS and checks and validate instructions coming from guest domains. Xen provides images that simulate commodity operating systems in ``porting'' ABIs to the hypervisor.

The process of configuring and booting a guest domain is more or less handy depending on the system used. A virtual machines monitor allows to control the machines up and running. A display or console can usually be opened from the host machine, and provided the guest VM is properly configured, users can log on it as if it were a real machine.\\

Xen is the virtualization technology used in the systems we present here. It is not the most easy to use but has gained very wide acceptance amongst Linux users: it offers a performance close to native speed, which used to be far better than its most important competitor VMWare before they also implemented para-virtualization. It is open, still more mature than KVM, and provides the most advanced features in its category.
\subsection{Benefits}
Platform virtualization allows a single host to bring up in different VMs different software and operating system distributions, and thus to satisfy the requirements of several applications at the same time. In addition, it is possible to flexibly and conveniently control the resource allocated to these VMs (e.g. CPU, memory). Finally, data isolation between VMs protect them against each other's failures and untrusted users, and it protects the native operating system against any threat from the VMs.

These benefits can be integrated into resource management systems for Platform virtualization systems usually present software interfaces or executables, thus allowing automation by other software.\\

However, grid resource management systems do benefit from platform virtualization only marginally. Major scientific grids do not yet provide users virtual platforms in production. Globus investigates the topic with Virtual Workspaces \cite{Keahey:Foster}. Tycoon, the system we present in the next section, has an experimental running setup with NorduGrid, developed mainly by Thomas Sandholm; and we are contributing, based on the work of Lars Rasmusson and Niklas Wirstrom from BalticGrid/SICS, to its integration with EGEE.
\section{Tycoon, dynamic sharing CPU capacity for usage optimization} \label{tycoon}
Tycoon is a resource management system that builds on virtualization to implement a specific resource management strategy: resource usage optimization via a bidding system and a global economy. In this aim, it benefits from the resource sharing capability of virtual machines, and in particular the ability of Xen to dynamically adjust the proportion of CPU cycles allocated to different VMs. In addition, as a system that brings up execution environments to remote users, it benefits from data isolation between VMs: \textsl{a priori} untrusted users can access VMs, install and run their own software without threatening each other nor the backing OS.\\

The goal is to optimize resource usage in a world-wide economy, by:
\begin{itemize}
\item enabling resource consumers to use foreign machines, and giving them an incentive to use the least loaded machine, and inside the machine, the minimum resource required;
\item enabling resource owners to safely open their resource to foreign users, and giving them an incentive to do so.
\end{itemize}
\subsection{System}
\begin{figure}
\centering
 \includegraphics[scale=.43]{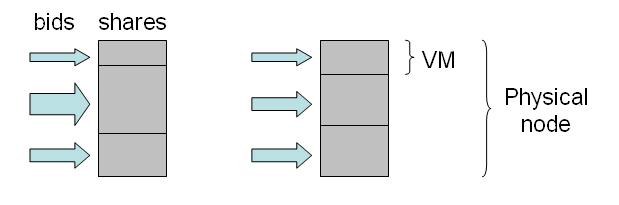}
\caption{This schema explains the relation between the bid and the resource share allocated to the bidder.}
\label{sharing}
\end{figure}

Tycoon's original idea is the creation of a global market for computing resources.
Anybody willing to use foreign resource needs a Tycoon client. Anybody willing to provide their own resource to foreign users and thus gain credits allowing to use foreign resource in turn, needs to install a Tycoon auctioneer on the machines he or she wants to involve in the process. The client is able to choose and connect to an appropriate auctioneer. The auctioneer then boots a VM on the host where it is installed, allocates a share of this host's resources to the VM (figure \ref{sharing}), and forwards the user's connection to this VM once it is booted.\\

This results in the distinction between two groups of Tycoon users: \emph{resource providers}, who run a Tycoon auctioneer on one or many of their own machines; and \emph{resource users}, who run a Tycoon client. These two groups widely overlap. This leads to balancing the utilization load over the whole grid of Tycoon-enabled machines: while a resource owner does not use the whole power of his machines, he gets credits from foreign users connecting to them; and at times when he needs more power than his own machines can offer he uses the accumulated credit to use more machines than he owns.\\

The load is also balanced inside one physical host by means of the auctioneer and bidding system: while the remote user is connected to his or her VM, the client can still communicate bidding adjustments to the auctioneer, and the auctioneer can in consequence dynamically adjust the share of CPU capacity allocated to each VM. The proportion of CPU capacity allocated to a particular VM inside a physical host is then the fraction of its user's bid amount per planned seconds of use out of the sum of bid amounts per seconds for VMs sitting on the same physical host. This mechanism is meant to ensure that CPU capacity is distributed according to its utility.\\

\begin{figure}
\centering
 \includegraphics[scale=.43]{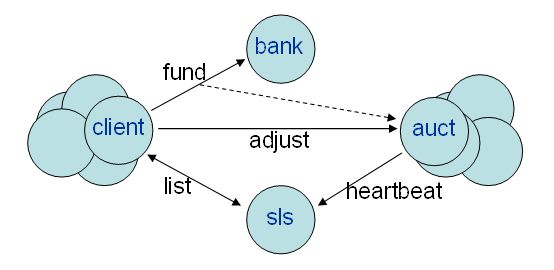}
\caption{The client, using the bank and Service Location Service, can either start a new bid on an appropriate host or adjust the amount of time a bid covers. The latter option results in adjusting the capacity allocated at a given time, without communicating with the bank.}
\label{architecture}
\end{figure}

A central \emph{bank} holds all the users and providers credits and proceeds with transactions when a client makes a bid and is granted access to a VM. A central \emph{service location server} keeps an up-to-date database of all available provider hosts for queries (figure \ref{architecture}). For example, the client can query a list ranked by increasing price of CPU capacity: the top host will be the one where the lower bid gives access to the highest CPU capacity. Clients run the \emph{best response algorithm} to bid automatically and thus relieve the user from the burden of constantly checking if the CPU capacity allocated to their VMs is sufficient.

\subsection{Tycoon and virtual resource management}
Tycoon belongs to the virtual resource management systems that provide an execution environment to remote users under the form of a VM. Among such systems, Tycoon shows particularly original and sophisticated management mechanism.
\emph{Virtual Workspaces}, developed at Argonne Labs, focuses more on compliance to grid standards (SOA). It presents a web-service interface that links sub-services together and allows administrators to run remote commands in order to manage filesystem images, VM booting, users, etc. \emph{Amazon Elastic Compute Clouds} (EC2) and \emph{Xenoservers} \cite{Hand:Harris} focus on letting users bring up their custom filesystem images using templates and diffs, and fast transfer technologies. In EC2, there is no fine-grained resource allocation like in Tycoon: the allocated VMs are of fixed size and characteristics, and users can only change how many of them they get. HP also experimented using Xen VMs to lower power consumption in data centers \cite{Kallahalla:Uysal}.\\

Tycoon is an example of how far we can get in the construction of a resource management mechanism and strategy based on virtualization. In next section we present a system we developed to ease research and development of such mechanisms.
\section{SmartDomains, enabling automation with SmartFrog} \label{smartdomains}
SmartDomains (SD), to contrast with Tycoon, does not focus on specific benefits of virtualization nor does it implement any specific mechanism for resource management. Rather it makes it possible to develop such mechanisms in a convenient way, based on a Java framework from HP Labs called SmartFrog.
\subsection{SmartFrog}
SmartFrog (SF) is a Java framework developed at HP Labs. It offers a rich description language and a deployment mechanism to configure, orchestrate, place and sustain SmartFrog applications made with Java classes to be distributed across many hosts. By SmartFrog application, we mean a running collection of synchronized components communicating via Java RMI. It is not necessarily an end-user application, but can be part of the infrastructure supporting it. In the case of SmartDomains, the SF application consists in the deployed pool of VMs.\\

In order to deploy a distributed application, when each component needed is available (in a Jar file stored locally or on a class-loading server), one has to write a description in SF language, and launch it with a simple local start command. A peer-to-peer layer of SmartFrog daemons runs on each machine potentially involved, in order to distribute and process the deployment and management tasks. To terminate the application it is again a matter of a simple local termination command.\\

To counterbalance the execution simplicity, the description language allows to define advanced setups so that they can be handled without human intervention at management time. To each SF component corresponds a component in the description, where configuration values can be defined. In addition, other components can be included, thus defining a parent-child relationship, where the parent handles the child's deployment, start and termination. The Java class of each component defines what is to be done within these phases.\\

Functionality, coded in the deploy, start and terminate methods, is separated from configuration, coupling and deployment processes dealt with by SmartFrog. Adding components then consists in coding the corresponding classes. Building the blocks is done in the description where components are easily associated. Any new functionality translated into a new component is then highly configurable: fields of the Java class can be presented as attributes of the corresponding component in the description. References to other components in the tree can be included in order to be resolved at run-time, thus enabling the remote invocation of a method from the referenced component.

The description language provides mechanisms to link an attribute name to a component or a value following a path up or down the tree, or to place an attribute likewise. The extension mechanism allows to name a new component that branches off another in overriding the values of its attributes or adding new attributes. One particular attribute, \emph{sfClass}, refers to the corresponding Java class for the component, and this is how the ``description space'' translates into building blocks from the ``code space''. A component name, \emph{sfConfig}, is reserved to the root of the tree, which is looked up by the parser when the start command is launched.
\subsection{SmartDomains components}
\begin{figure}
\centering
 \includegraphics[scale=.35]{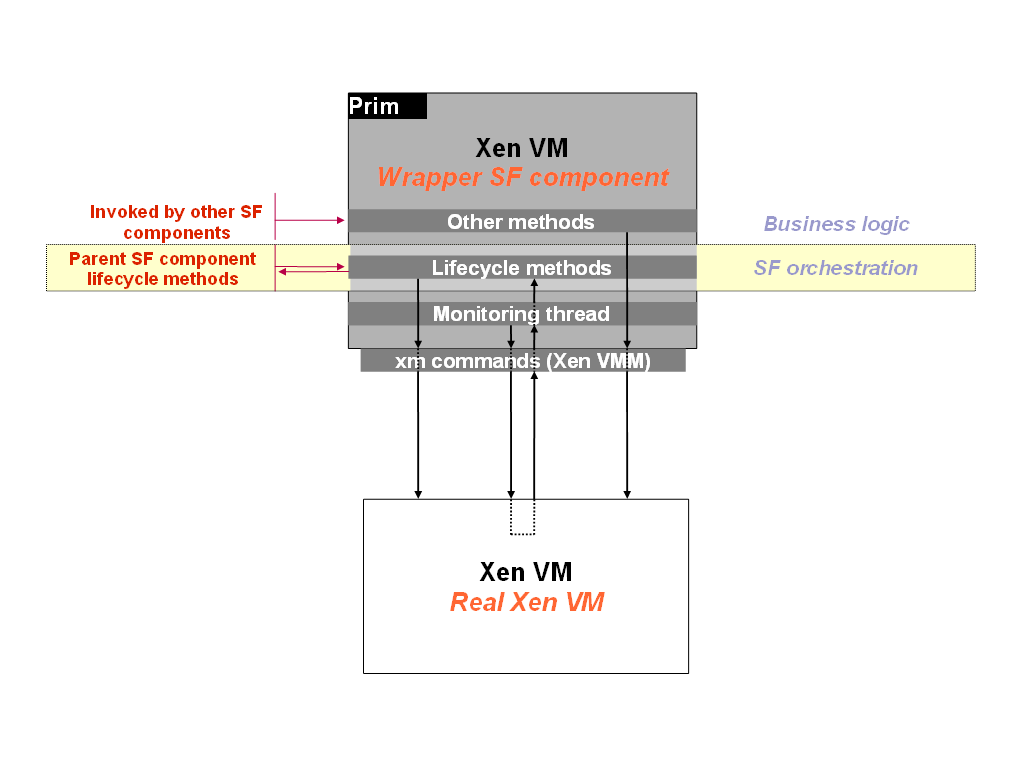}
\caption{A conceptual representation of the base SmartDomains components, their interaction with SmartFrog and with a VM via the Xen Virtual Machine Monitor (VMM) interface.}
\label{sdimage}
\end{figure}

SmartDomains is a set of SmartFrog components, the functionality of which consists of booting, monitoring and terminating VMs. The idea is to use the power of SmartFrog to configure, deploy and manage distributed pools of VMs (figure \ref{sdimage}). These base components represent running VMs and the way they are installed on the host filesystem. Building on them, resource management systems can be formed by plugging components that implement management functionality. Any of these components can then be reused, reconfigured, and associated in different manners in different descriptions. The configurability of virtual machines is not reduced by delegating their configuration to the SF descriptions. This was one of the main concerns during the implementation of the base components. As a consequence, translating Xen VMs into SmartFrog components allows to build complex systems involving execution environments, like Tycoon, without reducing \textsl{a priori} the configuration capabilities of these execution environments, nor enforcing any particular mechanism or strategy; so that any new building block could benefit to other systems. The ability to change the whole organization of a system without modifying the code is particularly useful in a research perspective where different strategies for managing virtual execution environments have to be evaluated.

\begin{itemize}
 \item{The storage back-end component} defines the way a VM is installed on the host filesystem. This involves logical volumes or loop-back devices management, OS image storage, de-compression at boot-time and compression at termination time, and such. Configuring this component means overriding default location values, setting disk space and swap size for the VM, etc.\\
 \item{The domain component} defines the running VM and the monitoring process. Since SmartFrog monitors the liveness of each component, it was interesting to monitor the liveness of the virtual machine from a component, so that high availability mechanisms could be implemented, even if only for code correctness and consistency. This is where runtime parameters of the VMs are defined as well as parameters not regarding the backing host filesystem itself. For example, the number of virtual CPUs, the allocated memory, and so on.
\end{itemize}
At deployment time, all these parameters are resolved by both components. When the storage back-end component starts, it prepares the image and the backing filesystem, and then it boots the virtual machine. When the domain component starts, it ``pings'' the VM to see when it is up and running, and then goes on to monitor liveness. When the VM dies, it cleanly removes it and notifies the parent component. The appropriate action upon this notification is defined while in the description by which parent component is used. Normally, the parent component is a ``Compound'' component, so that when the domain terminates, the storage back-end termination is also called, which brings the host filesystem back to the initial clean state.\\

SmartDomains is only at the beginning of the possibilities. Base components can still be expanded to cover more functionality. It does not yet control the CPU capacity allocation whereas Tycoon does. We are starting to implement management components and investigate which functionality we might insist on first: features for high availability or load balancing for example. We are looking for use cases that would need such features in order to drive their development in a usage-oriented process, just like we did for the base components. Indeed, thanks to the work of Olivier Pernet who has contributed to its development, our components have been used at CERN for gLite certification testbeds since the early stages of their development. SmartDomains brings up synchronized, distributed VMs as testbeds for gLite middleware quality assurance tasks. We benefited from a constant feedback from Andreas Unterkircher who drives the use case, and gLite testers experience with SmartDomains was simplified and improved by a web interface developed by Omer Khalid (vGrid).
\section{Conclusion}
Although virtualization has become a commodity, and we already see instances of systems managing virtual execution environments in commercial services, public scientific grids have not yet caught the wave, and we believe that many tracks still remain to be explored in this domain. We presented Tycoon, that implements a market-based mechanism to optimize resource usage on a global scale by balancing the load in time and space while prioritizing execution environments according to their utility. With other systems, experimentation addresses different needs like power consumption and cooling. In order to ease developments and experiments, we enabled the management of VMs via SmartFrog, a powerful framework that configures and deploys distributed systems in a peer-to-peer, component based model, eases modules reuse and architecture restructuring. With these illustration and development, we hope to foster research and collaboration, so that utility computing systems bringing up execution environments based on platform virtualization reach Grid data centers.

\end{document}